\documentclass[aps]{revtex4}
 
\usepackage{amsmath}
\usepackage{amssymb}
\usepackage{graphicx}

\newcommand{\beq}{\begin{equation}}
\newcommand{\eeq}{\end{equation}}
\newcommand{\ben}{\begin{eqnarray}}
\newcommand{\een}{\end{eqnarray}}
\newcommand{\benn}{\begin{eqnarray*}}
\newcommand{\eenn}{\end{eqnarray*}}

\newcommand{\qs}{q_{1s}^k}
\newcommand{\qf}{q_{1f}^k}

\newcommand{\pd}{\partial}
\newcommand{\rr}{\sqrt{r}}
\newcommand{\linv}{\mathcal{L}^{-1}}
\newcommand{\pc}{\mathcal{P}}
\newcommand{\apc}{\alpha_{\pc}}
\newcommand{\tl}{\tilde{\Lambda}}
\newcommand{\tp}{\tilde{\pc}}

\begin{document}

\title{Mode signature and stability  for a Hamiltonian model of electron temperature gradient turbulence}

\author{E. Tassi$^1$, P.J. Morrison$^2$}
\affiliation{$^1$ Centre de Physique Th\'eorique, CNRS -- Aix-Marseille Universit\'es, Campus de Luminy, case 907, F-13288 Marseille cedex 09, France \\ $^2$ Institute for Fusion Studies and Department of Physics, The University of Texas at Austin, Austin, TX 78712-1060, USA}
\date{\today}
\baselineskip 24 pt

\begin{abstract}
Stability properties and mode signature for equilibria of a model of  electron temperature gradient (ETG) driven turbulence are investigated  by Hamiltonian techniques. After deriving the infinite families of Casimir invariants, associated with the noncanonical Poisson bracket of the model, a sufficient condition for  stability is obtained by means of the Energy-Casimir method.  Mode signature is then investigated for linear motions about homogeneous equilibria. Depending on the sign of the equilibrium ``translated''  pressure gradient, stable equilibria can either be energy stable,  i.e.\ possess definite linearized perturbation energy (Hamiltonian), or spectrally stable with the existence of  negative energy modes (NEMs). The ETG instability is then shown to arise through a Kre\u{\i}n-type bifurcation, due to the merging of a positive and a negative energy mode, corresponding to two modified drift waves admitted by the system. The Hamiltonian of the linearized system is then explicitly transformed into normal form, which unambiguously  defines  mode signature. In particular, the fast mode turns out to always be a positive energy mode (PEM), whereas  the energy of the slow mode  can have either positive or negative sign. 
\end{abstract}
\maketitle

\section{Introduction}  \label{sec:intro}

An important issue for the stability of equilibria of continuous media concerns the existence of   {\it negative energy modes} (NEMs),  spectrally  stable   modes of oscillation of a  medium with  negative energy.    One reason NEMs are important is because equilibria with them, although linearly or spectrally stable, can be destabilized by arbitrarily small perturbations.  For example, if dissipation is added to the dynamics so as to remove energy from a NEM, then it can be proven that the mode becomes spectrally unstable.   On the  intuitive level,   dissipation  removes energy from  the  already negative energy of the mode, which makes it more negative and increases  the amplitude of mode.   In non-dissipative systems, NEMs  can become unstable with  the presence of  positive energy modes (PEMs)  through nonlinear coupling.  By this means the system can even develop finite-time singularities  while conserving the energy of the nonlinear system (see, e.g. \cite{Mor98} and many original references therein).

In plasma physics, the study of NEMs has a long tradition dating to the early work of Sturrock \cite{ST58} on streaming instabilities  and Greene and Coppi  \cite{Gre65} on magnetohydrodynamic (MHD) type dissipative instabilities in confinement systems.  NEMs have been studied in many plasma contexts; for example,   Vlasov-Maxwell dynamics \cite{Mor89,Mor90,Mor92,Cor92,Cor93,Cor97}, Maxwell drift-kinetic \cite{Thr96} theories, wave-wave interaction in the two-stream instability \cite{Kue95a,Kue95b,Las07}, magnetic reconnection \cite{Tas08}, ideal MHD in the presence of equilibrium flows \cite{Hir08,Hir08b,Ilg09}, magnetorotational instability \cite{Ilg07,Kha08},  and magnetosonic waves in the solar atmosphere \cite{Joa97}.

Because the concept of NEM is intimately related to that of energy, it was  proposed  in \cite{Mor89b} that the Hamiltonian framework  is  the natural one for defining and investigating this phenomenon, contrary to the usual practice in plasma physics.   Indeed, once the Hamiltonian structure of the model under consideration is known, an unambiguous definition of the energy of the system becomes available: the total energy corresponds to the Hamiltonian of the system and the energy of the linear dynamics must come in a natural way from the  second variation of this nonlinearly conserved quantity. Moreover, the normal form theory for linear Hamiltonian systems, provides a clear and systematic way for determining the signature of modes in the neighborhood of an equilibrium of the system. Indeed, for systems with discrete degrees of freedom, the Hamiltonian of the linearized system can always be cast, for stable modes, into the sum of Hamiltonians of decoupled harmonic oscillators, each of which possesses  a characteristic frequency and a {\it characteristic signature}.  Namely this signature, which, for each mode, can be positive or negative depending on whether the mode provides a positive or negative contribution to the total energy, provides a  systematic way to  identify PEMs or NEMs of the system. Finally, given the existence of the Hamiltonian structure,  energy-based methods, akin to `$\delta W$' of MHD, can be used to obtain sufficient conditions for stability of equilibria or to indicate the presence of NEMs. More precisely, an equilibrium has a NEM, if it is spectrally stable but the second variation of its free energy functional, evaluated at that equilibrium, has indefinite sign.

In this paper,  we investigate the presence of NEMs and stability properties of a reduced model for electron temperature gradient (ETG) instabilities, in the Hamiltonian framework. ETG turbulence has been considered as one of the mechanisms that enhances anomalous particle and electron thermal fluxes in tokamaks \cite{Hor88,Lee87}. The detection of NEMs, is therefore important in order to see what potentially unstable modes might lie dormant in the absence of dissipation. When destabilized by dissipation, such modes might enhance the anomalous transport across the confining magnetic field.  We note, however, that the methods applied for the ETG model, can be applied, in principle, to any ideal plasma model.\\
The model for ETG turbulence considered here, has been previously investigated in Refs. \cite{Gur04,Gur04b}, and in Ref.\cite{Gur04} it was described how this model possesses a  noncanonical Hamiltonian formulation,  with a Poisson bracket that turns out to be essentially identical to that for reduced MHD \cite{MorHaz84} (see also \cite{Mor82}).   In the present paper we first provide further information about the Hamiltonian structure of the model by deriving explicitly its families of Casimir invariants. We then determine sufficient conditions for  stability of generic equilibria by making use of the Energy-Casimir method. The investigation of the presence of NEMs is carried out for the case of homogeneous equilibria, for which we derive an explicit condition for the presence of NEMs. This condition warns us that NEMs are present if the value of the equilibrium pressure gradient lies in a given interval whose end points  depend on the perpendicular wave vector and the magnetic field curvature. In particular, the length of this interval shrinks to zero as the perpendicular wave number goes to infinity.  After obtaining the eigenvalues and eigenvectors of the system, we explicitly construct and  carry out the transformation that puts the Hamiltonian for the ETG model into normal form.

The paper is organized as follows. In Sec.\ref{sec:hametg} we review the model and its Hamiltonian formulation, and  then derive the Casimir invariants. In Sec.\ref{sec:fstab} we apply the Energy-Casimir method and obtain conditions for energy stability. In Sec.\ref{sec:nems}, after reviewing the theory of mode signature for linear Hamiltonian systems we apply  it to the ETG model.  Explicit conditions for the existence of NEMs, their relationship to energy stability  and spectral stability conditions coming from the dispersion relation  are derived, and the normal form transformation is explicitly obtained.  Finally, we conclude in Sec.\ref{sec:concl}.

\section{Hamiltonian structure of the ETG model} 
\label{sec:hametg}

The non-dissipative ETG driven turbulence model of  \cite{Gur04,Gur04b} is given by
\ben 
\frac{\pd }{\pd t}(1-\nabla^2)\phi &=& [\phi,\nabla^2\phi+x]+\left[\frac{p}{\rr},\rr x\right]\,, 
\label{e1} \\
\frac{\pd }{\pd t}\frac{p}{\rr} &=& \left[\frac{p}{\rr},\phi\right]+[\rr x,\phi]\,, 
 \label{e2}
\een
where two-dimensional slab geometry is assumed so that the field variables, $\phi$ the stream function and $p$  the pressure,  are functions of the Cartesian coordinates $(x,y)$.  The quantity $[f,g]=\partial f /\partial x \partial g /\partial y - \partial f /\partial y \partial g/\partial x$ is the  canonical Poisson bracket.   Equations (\ref{e1}) and (\ref{e2}) are written in the normalized form described in \cite{Gur04}, where the constant parameter $r$ is defined as
\beq
r=\frac{L_n^2}{L_B L_P},
\eeq
with
$L_n$, $L_B$ and $L_P$ being  the characteristic length scales of variation of the background density, magnetic,  and electron pressure fields, respectively. The parameter $r$ is thus related to the mechanism providing the drive for the ETG instability.  In particular, in the limit $r \rightarrow 0$, which is attained if one flattens the electron pressure gradient, the linear dispersion relation indicates  that ETG modes degenerate into marginally stable drift waves \cite{Gur04b}.

In   \cite{Gur04}, the authors showed that the system (\ref{e1})-(\ref{e2}) possesses a Hamiltonian structure in terms of a noncanonical Poisson bracket. This means (see, e.g. \cite{Mor98,Mor82}) that the system can be cast in the form
\beq  \label{sysham}
\frac{\partial \chi^i}{\partial t} = \{\chi^i , H \}, \qquad i=1, \cdots ,n
\eeq
with $\chi^i(\mathbf{x},t)$ indicating a suitable set of $n$ field variables (with $n=2$ in our case) and $H[\chi^1, \cdots ,\chi^n]$ a Hamiltonian functional that is conserved by the dynamics. The Poisson bracket $\{ ,\}$ appearing in (\ref{sysham}) is an antisymmetric bilinear binary operator satisfying the Leibniz rule and the Jacobi identity. For the model of (\ref{e1}) and (\ref{e2}),  it was shown \cite{Gur04} that, with the choice $\chi^1 = \phi$, $\chi^2 = p$, the Hamiltonian of the system is
\beq
H[\phi,p]=\frac{1}{2}\int d^2 x \left(\phi^2 +|\nabla\phi|^2-\frac{p^2}{r}\right)
\label{ham}
\eeq
and the Poisson bracket is 
\beq \label{pb0}
\{F,G\}=-\int d^2 x (\phi-\nabla^2\phi-x)[\mathcal{L}^{-1}F_{\phi},
\mathcal{L}^{-1}G_{\phi}]+\left(p+r x\right)([\linv F_{\phi},G_p]+[F_p,\linv G_{\phi}]),
\eeq
where the operator $\mathcal{L}$, and its inverse $\linv$ are formally defined so that $\mathcal{L}f=f-\nabla^2 f$, and $\linv \mathcal{L} f = \mathcal{L} \linv f =f$, for a generic field $f$. The subscripts on $F$ and $G$ in (\ref{pb0}) denote functional derivatives with respect to the fields $\phi$ or $p$.

Noncanonical Poisson brackets such as (\ref{pb0}) are characterized by the presence of so called Casimir invariants (see, e.g. Ref. \cite{Mor98}), due to  degeneracy in the cosymplectic operator of the bracket. More precisely, a Casimir invariant  of a Poisson bracket is a functional $C(\chi^1, \cdots , \chi^n)$ that satisfies
\beq  \label{casdef}
\{C, F \} =0
\eeq
for any functional $F$ of the field variables. Because they commute in particular with any $H$, Casimir functionals are preserved during the dynamics. In order to derive the Casimir invariants of the ETG model, it is convenient to introduce the variables
\beq
 \pc=\frac{p}{\rr}+\rr x, \qquad \lambda=\phi-\nabla^2\phi-x,
 \eeq
 which correspond to a ``translated" pressure and to a variable  analogous to  the potential vorticity of the Charney-Hasegawa-Mima equation \cite{Wei83,CTM10}, respectively.   In terms of these variables the model equations read
 \ben
 \frac{\pd}{\pd t}\lambda &=& -[\linv( \lambda + x),\lambda]+[\pc,\rr x],\\
 \frac{\pd}{\pd t}\pc &=& [\pc,\linv(\lambda+x)],
 \een
 whereas the Hamiltonian and the bracket become 
 \ben
 H(\lambda,\pc)=\frac{1}{2}\int d^2 x \left( (\lambda+x)\linv (\lambda +x)-\pc^2 + 2 \rr \pc x\right),\\
 \{F,G\}=-\int d^2 x \lambda[F_{\lambda},G_{\lambda}]+\pc([F_{\lambda},G_{\pc}]+[F_{\pc},G_{\lambda}])\,,
  \label{br2}
 \een
 where the bracket is seen to be identical to that for reduced MHD as first given in \cite{MorHaz84}.

Applying (\ref{casdef}) we deduce that the equations determining the Casimir invariants for our system are
\begin{equation} \label{eqCam}
\begin{split}
&[C_{\lambda},\lambda]+[C_{\pc},\pc]=0,\\
&[C_{\lambda},\pc]=0.
\end{split}
\end{equation}
By solving (\ref{eqCam}) we see that the system admits two independent infinite families of Casimirs: 
\ben \label{casim}
C_1=\int d^2 x\,  \mathcal{H}(\pc), \qquad C_2=\int d^2 x\,  \lambda \mathcal{F}(\pc),
\een
with $\mathcal{H}$ and $\mathcal{F}$ arbitrary functions. The dynamics described by the inviscid ETG model is then subject to an infinite number of constraints imposed by the conservation of the Casimir invariants (\ref{casim}). For instance, as a consequence of the conservation of $C_2$,   integrals of the potential vorticity $\lambda$ over regions bounded by contour lines of $\pc$ will be conserved during the dynamics (see \cite{Mor87}).

Notice that the constant of motion $\mathcal{I}$  found in \cite{Gur04} is given by
\begin{eqnarray}
\mathcal{I}&=&\int d^2 x \left(\frac{p^2}{r}+2(\phi-\nabla^2\phi)p\right)\nonumber\\
&=&
2\rr\int d^2x \lambda \pc +\int d^2 x \pc^2-2r \int d^2 x x\left(\phi-\nabla^2\phi-\frac{x}{2}\right)\,,
\end{eqnarray}
which is a linear combination of two particular Casimirs  of (\ref{casim}) with the realization that the time derivative  of 
\beq
 -2r \int d^2 x x\left(\phi-\nabla^2\phi-\frac{x}{2}\right)
 \eeq
vanishes  if $\pd \phi/\pd x$ and $\pd \phi/\pd y$ vanish or are periodic at the boundaries.
 Therefore,   $\mathcal{I}$ amounts to  a single special case of the general families of (\ref{casim}). 
 
\section{Energy stability}   \label{sec:fstab}

The Hamiltonian formalism provides a systematic procedure for implementing the Energy-Casimir method for investigating stability of equilibria (see, e.g.\ \cite{Mor86,Mor98,Mar99}), a stability method that originated in plasma physics in \cite{Kru58} that  has  often been referred to as nonlinear stability.  This method has been adopted in many works;  for example, in the context of fluid models for plasmas in  \cite{Haz84,Hol85,Kha05,Tas09}.  It provides sufficient conditions for stability by taking the second variation of the free energy,  the Hamiltonian plus Casimir invariants,  and extracting conditions  that are necessary for definiteness.     Because the method is based on nonlinear constants of motion, the  stability conditions obtained are stronger than conditions that emerge from dispersion relations, i.e.\ spectral stability conditions,  that follow entirely from the linear equations of motion.  Indeed this second variation stability, which we will refer to simply  as {\it energy stability},  implies linear stability \cite{Mor98,Bok02}, but the converse is not true.   In some works energy stability is  called formal stability when an additional  convexity estimate is not provided.   Usually these estimates are rather trivial and even when they are provided  they are only a small part of a mathematically rigorous  stability proof -- for this reason we eschew this terminology.

For noncanonical Hamiltonian systems,  equilibrium solutions can be found by solving the equations that result from extremizing the free energy functional.   For our system the free energy functional is given by  $F=H+C_1+C_2$ (which is not to be confused  with the generic functional $F$ of our Poisson brackets), with $H$  given by (\ref{br2}) and $C_{1,2}$ by (\ref{casim}).    For convenience we introduce the new variables according to the transformation
 \beq  \label{var2}
 \Lambda=\lambda +x, \qquad \bar{\pc}=\pc 
 \eeq
and then drop the bar on $\bar{\pc}$ in the following.    In terms of these variables the  Hamiltonian and the bracket become 
 \benn
 H(\Lambda,\pc)&=&\frac{1}{2}\int d^2 x \left( \Lambda\linv \Lambda -\pc^2 + 2 \rr \pc x\right)\,,
\\
 \{F,G\}&=&\int d^2 x (x-\Lambda)[F_{\Lambda},G_{\Lambda}]-\pc([F_{\Lambda},G_{\pc}]+[F_{\pc},G_{\Lambda}]). \label{br3}
 \eenn

The free energy functional is then explicitly given by
\beq
F(\Lambda, \pc)=\frac{1}{2}\int d^2 x \left( \Lambda\linv \Lambda -\pc^2 + 2 \rr \pc x\right)+\int d^2 x \mathcal{H}(\pc)+\int d^2 x (\Lambda -x) \mathcal{F}(\pc)\,,
\eeq
and the equilibrium equations, obtained from setting the first variation $\delta F$ equal to zero, are
\ben \label{equil}
F_{\Lambda} &=& \linv \Lambda +\mathcal{F}(\pc)=0, \label{equil1}\\
F_{\pc} &=& -\pc + \rr x +\mathcal{H}'(\pc)+(\Lambda -x)\mathcal{F}'(\pc)=0, 
\label{equil2}
\een
where the prime denotes derivative with respect to the argument of the function.
Due to the presence of the arbitrary functions in the Casimirs, such equilibrium equations possess free functions. Specifying these selects from a class of equilibrium solutions. In particular,  choosing $\mathcal{F}$ corresponds to fixing the relation between the equilibrium stream function $\phi_{eq}=\linv \Lambda_{eq}$ and the equilibrium translated pressure $\pc_{eq}$.

As indicated above, an equilibrium solution of (\ref{equil1})-(\ref{equil2}) is energy stable (and therefore linearly) stable, if the second variation of $F$, evaluated at that equilibrium, has a definite sign. In terms of the variables $\phi$ and $\pc$, the second variation of $F$ reads
 \ben
 \delta^2 F=\int d^2 x \Big[ (1-\mathcal{F}'(\pc))|\mathcal{L}^{-1}\delta \Lambda|^2 +(1-2\mathcal{F}'(\pc))|\mathcal{L}^{-1}\nabla\delta\Lambda|^2+
 \mathcal{F}'(\pc)(\delta \Lambda+\delta \pc)^2 
 \nonumber\\
 -  \mathcal{F}'(\pc)(\mathcal{L}^{-1}\nabla^2 \delta\Lambda)^2 
+\Big(\mathcal{H}''(\pc)-1
+(\Lambda-x)\mathcal{F}''(\pc)-\mathcal{F}'(\pc)\Big)|\delta\pc|^2 \Big].
\label{d2F}
 \een
From (\ref{d2F}) one immediately obtains   sufficient conditions  for positive definiteness of $\delta^2 F$ in the case of no flow:
   \ben    
  \mathcal{F}(\pc_{eq})\equiv 0 \qquad {\rm and}\qquad 
\mathcal{H}''(\pc_{eq})>1 \,.
\label{forstab1}
  \een
{}From (\ref{equil1}), $\mathcal{F}(\pc_{eq})\equiv 0$ implies no flow, while $\mathcal{H}''(\pc_{eq})>1$  gives a condition on the pressure profile.  Generally speaking, the situation with flow, when $\mathcal{F}(\pc_{eq})\not\equiv 0$,  is expected to have NEMs \cite{Mor98}.  However, this case is more complicated to analyze,  with the Poincar\'e inequality often being of use, but we will not pursue it further here.


\section{Negative energy modes}  \label{sec:nems}

In this section, we first review the theory of NEMs in the finite degree-of-freedom Hamiltonian context \cite{KJ80}, since it applies directly to finite systems with discrete spectra (e.g.\ \cite{Kue95a}).  The situation for continuous spectra is  more complicated   \cite{Mor92,Mor00,HagMor10,Hir10} and will not be considered here.   Subsequently, after carrying out a spectral stability analysis  
of the system linearized around homogeneous equilibria with no flow, we  make use of the Hamiltonian formalism in order to
detect the presence of NEMs among  the stable modes of the linearized system.  Finally we carry out 
the explicit transformation that casts the corresponding Hamiltonian into normal form.


\subsection{Review of mode signature and normal forms for linear Hamiltonian systems}  \label{ssec:nem}

A real canonical Hamiltonian linear system with $N$ degrees of freedom is generated by  the canonical Poisson bracket
\beq
\{f,g\}= \frac{\partial f}{\partial q^i}\frac{\partial g}{\partial p_i}-\frac{\partial f}{\partial p_i}\frac{\partial g}{\partial q^i}\,,
\eeq
and a quadratic Hamiltonian
\beq
H_L=\frac{1}{2} A_{ij}z^i z^j,
\label{HL}
\eeq
where $z=(q_1,...,q_N,p_1,...,p_N)$,  $A_{ij}$ are the elements of a $2N\times2N$ matrix with constant coefficients, and repeated sum notation is assumed with $i,j=1,2,\dots, N$. The resulting equations of motion can then be compactly written as
\begin{equation}  \label{lingen}
\dot{z}=J_c A z,
\end{equation}
where 
\beq
J_c=\left( {\begin{array}{cc}
 0_N & -I_N  \\
 I_N & 0_N  \\
 \end{array} } \right)
\eeq
is the $2N\times 2N$ canonical symplectic matrix (cosymplectic form). Assuming 
\beq
z=\tilde{z}\mathrm{e}^{i\omega t}+ \tilde{z}^{*}\mathrm{e}^{-i\omega t},
\eeq
with $^*$ indicating complex conjugate, (\ref{lingen}) yields the eigenvalue problem  
\beq   \label{linfou}
i\omega_{\alpha}z_{\alpha}=J_c A z_{\alpha}, \qquad \alpha=1,\cdots ,N,
\eeq
where we have dropped the tilde on the eigenvectors and have added an eigenvalue label $\alpha$.   In what follows we will assume distinct eigenvalues, precluding the existence of nontrivial Jordan form and possible secular growth in time.  We also assume that the eigenvalues $\omega_{\alpha}$ are real, which is the case of interest for detecting mode signature.   Because our dynamical variables are real, the remaining $N$ eigenvalues are given by $\omega_{-\alpha}=-\omega_{\alpha}$ and the corresponding eigenvectors are $z_{\alpha}=z_{\alpha}^*$.  Defining $\Omega:=J_c^{-1}$, the symplectic two-form, we construct the quantity
\beq
h(\alpha,\beta):= i\omega_{\alpha}z_{\beta}^T\Omega z_{\alpha}=z_{\beta}^T A z_{\alpha}\,,
\eeq
where $T$ denotes transpose. 

It can be easily shown that the property $h(\alpha,\beta)-h(\beta,\alpha)=0$ holds. Then, from this 
relation and the antisymmetry of $\Omega$, it follows that 
\beq
h(\alpha,\beta)=0, \qquad \mbox{if} \quad \beta \neq -\alpha.
\eeq
On the other hand,  %
\beq
h(-\alpha,\alpha)=z_{-\alpha}^T A z_{\alpha}={z_{\alpha}^*}^T A z_{\alpha}=i\omega_{\alpha}{z_{\alpha}^*}^T \Omega z_{\alpha},
\eeq
is clearly the  {\it energy} (Hamiltonian $H_L$) of  the mode $(z_{\alpha},\omega_{\alpha};z_{\alpha}^*,-\omega_{\alpha})$. Evidently, ${z_{\alpha}^*}^T \Omega z_{\alpha}$ is a purely imaginary number, and a normalization constant for  the eigenvectors can be chosen in such a way that
\beq \label{lagr}
{z_{\alpha}^*}^T \Omega z_{\alpha}=\pm 2i,
\eeq
with the sign, an invariant,  depending on the specific mode under consideration. Note that the left-hand side of (\ref{lagr}) represents the Lagrange bracket (symplectic two-form) of  $z_{\alpha}^*$ and $z_{\alpha}$.  If $z_{\alpha}$ is an eigenvector, associated with  a positive eigenvalue $\omega_{\alpha}$, and 
\beq
{z_{\alpha}^*}^T \Omega z_{\alpha}=-2i,
\eeq
then $(z_{\alpha},\omega_{\alpha};z_{\alpha}^*,-\omega_{\alpha})$ corresponds to a  {\it positive energy mode}, otherwise it is a {\it negative energy mode}. This can be easily seen by observing that, in the case of a PEM, the corresponding energy is given by
\beq
h(-\alpha,\alpha)=i \omega_{\alpha}{z_{\alpha}^*}^T \Omega z_{\alpha}=2\omega_{\alpha}>0.
\eeq
Note that, although here we carried out an analysis with canonical coordinates, Sylvester's theorem guarantees that the signature of a mode (i.e., whether it is a PEM or a NEM), does not depend on the choice of the coordinate system.

The distinction between positive and negative energy modes becomes even more transparent when we are reminded that, for stable modes there exists \cite{Mor89,Kue95a,KJ80} a canonical transformation $T:(Q_1, \cdots , Q_N,P_1,\cdots , P_N) \rightarrow (q_1, \cdots q_N, p_1, \cdots p_N)$, that casts the  quadratic Hamiltonian of (\ref{HL}) into the following {\it normal form}:
\beq \label{nformgen}
H_L=\frac{1}{2}\sum_{\alpha=1}^N \sigma_{\alpha}\, \omega_{\alpha}  (P_{\alpha}^2 + Q_{\alpha}^2)\,, 
\eeq
where  $\omega_{\alpha}$ represents the  positive eigenvalues of the linearized system, whereas $\sigma_i \in \{-1,1\}$ is the signature of the mode. 

If the system contains unstable modes, then they have a different normal form.  However, if the Hamiltonian is restricted to the stable modes, then it can be written as the Hamiltonian for a system of $N$ harmonic oscillators with different frequencies. The modes for which $\sigma=-1$,   which give a negative contribution to the total energy, correspond to the NEMs, while those corresponding to $\sigma_i=1$ are, of course, PEMs.

Once the eigenvalues and eigenvectors of the system are known, the procedure for constructing the map $T$ is algorithmic. First one needs to select, among the $2N$ eigenvectors of the system, the $N$ eigenvectors $z_{\alpha}$ that satisfy 
\beq
{z_{\alpha}^*}^T\Omega z_{\alpha}=-2i, \qquad \alpha=1,\cdots,N.
\eeq
Then the $2N\times 2N$ matrix that defines the  transformation is given by
\beq
T= {\rm col}\, (
 \mathrm{Re} \, z_1,   \mathrm{Re}\, z_2\, ... \, \mathrm{Re} \, z_N,  \mathrm{Im}\,  z_1, \mathrm{Im}\,  z_2, \dots,  \mathrm{Im}\,  z_N
)\,,
\eeq
which is the matrix with columns given by $ \mathrm{Re}\,  z_1$ etc.  It can be shown that the transformation constructed in this way is canonical and indeed provides  the desired diagonalization.


\subsection{Mode signature and stability for the ETG model}  
\label{ssec:appetg}

Now consider a special case of the no-flow equilibria of Sec.~\ref{sec:fstab}, the   homogeneous equilibria:
 \beq \label{equili}
 \Lambda_{eq}=\linv \Lambda_{eq}=0, \qquad \pc_{eq}=\alpha_{\pc} x\,,
 \eeq
where $\alpha_{\pc}$ is a constant. The equilibrium solution (\ref{equili}) corresponds to the choices
 \beq  \label{chcas}
 \mathcal{F}(\pc)=0, \qquad \mathcal{H}(\pc)=\frac{1}{2}\left(1-\frac{\rr}{\alpha_{\pc}}\right)\pc^2,
 \eeq
for the Casimir functions that appear in  equations (\ref{equil1}) and (\ref{equil2}).

Linearizing the model equations around this equilibrium gives the system
\benn  \label{linsyst}
\dot{\tilde{\Lambda}}&=&-\frac{\pd }{\pd y}\linv \tilde{\Lambda}-\rr\frac{\pd }{\pd y}\tilde{\pc},\nonumber\\
\dot{\tilde{\pc}}&=&\apc\frac{\pd }{\pd y}\linv\tilde{\Lambda}\,.
\eenn
Expanding the perturbations as Fourier series, as follows: 
\beq
 \Lambda=\tilde{\Lambda}=\sum_{\mathbf{k}=-\infty}^{+\infty} \tilde{\Lambda}_{\mathbf{k}}(t)\mathrm{e}^{-i \mathbf{k}\cdot\mathbf{x}}, \qquad  \pc=\alpha_{\pc}x+\tilde{\pc}=\alpha_{\pc}x+\sum_{\mathbf{k}=-\infty}^{+\infty} \tilde{\pc}_{\mathbf{k}}(t)\mathrm{e}^{-i \mathbf{k}\cdot\mathbf{x}},
\eeq
yields the amplitude equations
\begin{eqnarray}
\dot{\tl}_{\mathbf{k}}&=&i\frac{k_y}{1+k_{\perp}^2}\tl_{\mathbf{k}}+i\rr k_y\tilde{\pc}_{\mathbf{k}}, 
\label{link1} 
\\
\dot{\tilde{\pc_{\mathbf{k}}}}&=&-i\apc\frac{k_y}{1+k_{\perp}^2}\tl_{\mathbf{k}}\,,  
\label{link2}
\end{eqnarray}
whence, the dispersion relation for modes of the form $\mathrm{e}^{i(\omega t -\mathbf{k}\cdot\mathbf{x})}$ is  obtained, 
\beq
\omega^2 -\frac{k_y}{1+k_{\perp}^2}\omega+\apc \frac{\rr \, k_y^2}{1+k_{\perp}^2}=0.
\eeq
This expression is in agreement with that obtained in Ref.\cite{Gur04b}. The eigenvalues correspond to a slow and a fast mode, and are given by
\ben \label{eigv}
\omega_s^k=\frac{k}{2(1+k_{\perp}^2)}(1-\sqrt{1-4 (1+k_{\perp}^2)\apc\rr}), \label{eigv1}\\
\omega_f^k=\frac{k}{2(1+k_{\perp}^2)}(1+\sqrt{1-4 (1+k_{\perp}^2)\apc\rr}), \label{eigv2}
\een
where we have set $k_y=k$. The corresponding eigenvectors are 
\beq \label{eigen}
\tl_{s,f}^{k}=-\omega_{s,f}^k\frac{1+k_{\perp}^2}{\apc k}\tp_{s,f}^{k}=-\frac{1}{2\apc}(1\pm\sqrt{1-4(1+k_{\perp}^2)\apc \rr})\tp_{s,f}^{k},
\eeq
for $k>0$. The system also possesses the eigenvalues $\omega_{-s,-f}^k=-\omega_{s,f}^k$, whose eigenvectors are the complex conjugates of those of (\ref{eigen}).

From (\ref{eigv1})-(\ref{eigv2}) we obtain a necessary and sufficient condition for  spectral stability, viz.
\beq
1-4(1+k_{\perp}^2)\apc \rr>0 \quad \Rightarrow \quad \apc < \frac{1}{4(1+k_{\perp}^2) \rr}.
\eeq
Therefore, if the electron pressure or magnetic field gradients are such that $\apc<0$, such equilibria are always spectrally stable $\forall\   k$.   If $\apc>0$, on the other hand, the equilibrium will be stable only for ${\bf k}$ such that $0< \apc < 1/4(1+k_{\perp}^2)\rr$ is satisfied. In other words, there will always be instability for sufficiently large $k_{\perp}$.

The linearized system of (\ref{link1}) and (\ref{link2}) inherits a 
Hamiltonian formulation from the nonlinear system, one that can be written in the framework described in Sec.~\ref{ssec:nem}.  We can then take advantage of this fact  in order to see whether NEMs are present in the system and to cast the Hamiltonian into its normal form.

By using the relation (see, e.g., \cite{Gar71,Mor80,Kue95a})
\beq
\frac{\delta F}{\delta \tilde{\Lambda}}=\sum_{k=-\infty}^{k=+\infty}\left(\frac{\delta F}{\delta \tilde{\Lambda}}\right)_k \mathrm{e}^{-iky}=\frac{1}{2\pi}\sum_{k=-\infty}^{+\infty}\frac{\partial \bar{F}}{\partial \tilde{\Lambda}_{-k}}\mathrm{e}^{-iky},
\eeq
where $F(\tilde{\Lambda})=\bar{F}(\tilde{\Lambda}_{k})$,
it can be shown that the Hamiltonian structure of  
(\ref{link1}) and (\ref{link2}), is given by the bracket 
\beq  \label{brk}
\{F,G\}=\sum_{k=1}^{+\infty}\frac{ik}{2\pi}\left[\left(\frac{\pd F}{\pd \tl_k}\frac{\pd G}{\pd \tl_{-k}}-\frac{\pd F}{\pd \tl_{-k}}\frac{\pd G}{\pd \tl_{k}}\right)-\apc^2\left(\frac{\pd F}{\pd \tl_k}\frac{\pd G}{\pd \tp_{-k}}+\frac{\pd F}{\pd \tp_k}\frac{\pd G}{\pd \tl_{-k}}-\frac{\pd F}{\pd \tp_{-k}}\frac{\pd G}{\pd \tl_{k}}-\frac{\pd F}{\pd \tl_{-k}}\frac{\pd G}{\pd \tp_k}\right)\right].
\eeq
and the linear Hamiltonian, which is proportional to $\delta^2 F$,  
\beq  \label{hamk}
H_L=\sum_{k=1}^{+\infty}H_L^k=2\pi\sum_{k=1}^{+\infty}\left( \frac{|\tl_{k}|^2}{1+k_{\perp}^2}-\frac{\rr}{\apc}|\tp_k|^2\right),
\eeq
where we have suppressed the sum on $k_x$ (note that $k_x$ only appears in the combination $k_{\perp}^2=k_x^2 + k^2$).
Although not canonical, this formulation, in principle is sufficient in order to detect the presence of NEMs for the equilibrium under consideration. Indeed, as already done for the four-field model of \cite{Tas08}, we can make use of the property that NEMs can change their signature only if they become unstable through a ``Kre\u{\i}n bifurcation'' \cite{Mor98}, or if the corresponding eigenvalues go through  zero frequency. It is then sufficient to identify a NEM in a particular limit, and we are then  guaranteed that its signature does not change as long as one of the two above mentioned phenomena does not occur. Sylvester's theorem also guarantees  that the signature is independent on the choice of the coordinates we make. If we fix a wave vector $k$,  then  we can first evaluate the energy associated with the  corresponding mode in the $(\tl_k,\tl_{-k},\tp_k,\tp_{-k})$ coordinates  by inserting eigenvalues and eigenvectors associated to $k$ in the expression for $H_L^k$. This results in 
\ben \label{flk}
{H_L}_{s,f}^k=2\pi\left(\frac{1 -4 \rr \apc 
- 2 k_{\perp}^2\apc\rr \pm \sqrt{1-4(1+k_{\perp}^2)\apc \rr}}{2 \apc^2}\right)|\tp_{s,f}^k|^2\,,
\een
which is the energy of the slow and the fast modes of  wave number $k$ (summed over $k_x$). In order to identify PEMs and NEMs,  it is sufficient to consider the limit $k_{\perp}\rightarrow 0$, which yields
\ben \label{fek0}
{H_L}_{s,f}^{k,k_{\perp=0}}=2\pi\left(\frac{1 -4 \rr \apc  \pm \sqrt{1-4\apc \rr}}{2 \apc^2}\right)|\tp_{s,f}^k|^2.
\een
We can then see that, for the fast mode (corresponding to the $+$ sign in (\ref{fek0})), ${H_L}_f^{k,k_{\perp=0}}$ is positive, and therefore a PEM. Also, it will remain a PEM  as parameters are varied in a continuous way,  until  the instability threshold is  reached. For the slow mode, two cases exist for finite $\apc$. If $\apc<0$, then ${H_L}_s^{k,k_{\perp=0}}>0$ and, again, we have a PEM. If $0<\apc<1/(4\rr)$, on the other hand, the slow mode is a NEM.

 The above mentioned instability, occurring at large $k_{\perp}$, for $\apc>0$, indicates a Kre\u{\i}n bifurcation, which is one of the possible types of bifurcations occurring in Hamiltonian systems. 
\begin{figure}
\centering
\includegraphics[width=6cm]{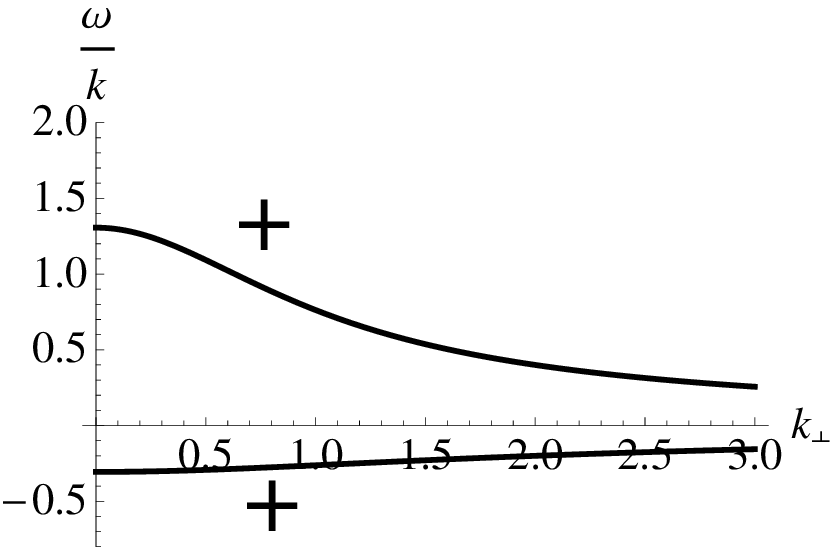}
\hspace{1cm}
\includegraphics[width=6cm]{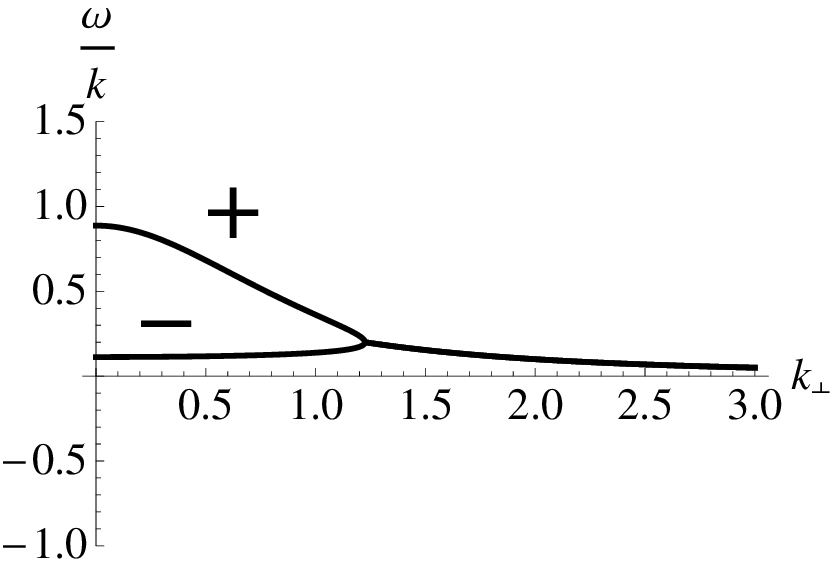}
\caption{Depiction of two possible mode signature situations for ETG modes,  depending on the sign of $\apc$. The plot on the left refers to the case of negative $\apc$ ($\apc=-0.3, \sqrt{r}=0.2$). In this case both modes are PEMs and the system is energy  stable. In the plot on the right,  $\apc$ is positive ($\apc=0.5, \sqrt{r}=0.2$). For $0<k_{\perp}<1.22$,  the equilibrium is stable,  but the slow mode is a NEM. A Kre\u{\i}n bifurcation occurs at $k_{\perp}=1.22$.}  
\label{fig1}
\end{figure}
These situations are illustrated in Fig.~\ref{fig1}. When $\omega_s^k$ is real and negative, that is for $\apc <0$, both the slow and the fast modes are PEMs and the two branches correspond to two dispersive waves with opposite sign. These waves correspond to inviscid drift waves  modified by the presence of the ETG and magnetic field curvature. In the limit $r\rightarrow 0$, corresponding to vanishing ETG, the two modes degenerate into a Hasegawa-Mima drift wave. When $\omega_s^k$ is real and positive, i.e.\  for $0<\apc< 1/4(1+k_{\perp}^2)\rr$, both modes are still stable but the slow mode is now a NEM. Comparing the two plots of Fig.\ref{fig1}, one observes that $\omega_s^k$ went from negative to positive, i.e.\  it crossed through zero frequency, while changing from a PEM to a NEM. For the parameters chosen for the figure, the instability threshold, due to the presence of ETG, occurs at $k_{\perp}\approx1.22$. For perpendicular wave numbers above this value, the equilibrium is indeed unstable. The transition of the two real eigenvalues into a complex conjugate pair, occurring at $k_{\perp}\approx 1.22$, is an example of  a Kre\u{\i}n bifurcation. Note that, as predicted by Kre\u{\i}n's theorem (see, e.g., Ref. \cite{Mor98}), if a Kre\u{\i}n bifurcation between two eigenvalues occurs, one of them must be a NEM. This is indeed our case.

If we consider now the  stability condition (\ref{forstab1}) for no-flow equilibria, together with (\ref{chcas}), we find that the homogeneous equilibrium (\ref{equili}) is energy stable if $\mathcal{H}''(\pc_{eq})>1$, which is equivalent to $\apc <0$. Indeed,  if this condition is satisfied, the equilibrium is stable, with no NEMs. If we now use the actual pressure $p$, as a variable, and consider the corresponding homogeneous equilibrium $p_{eq}=a_p x=\rr(\apc-\rr)x$, we can reformulate our results in the following way. If the equilibrium pressure gradient is such that $a_p < -r$ (i.e., $\apc <0$), then the system is energy stable for every ${\bf k}$. If, on the other hand, $-r < a_p < -r + 1/4(1+k_{\perp}^2)$, (i.e. $0<\apc<1/(4(1+k_{\perp}^2)\rr$) then the system becomes unstable through a Kre\u{\i}n bifurcation at a critical $k_{\perp}$, for fixed $r$. For $k_{\perp}$ below this  critical value, the equilibrium is spectrally stable but not energy stable. Indeed, the slow mode in this case is a NEM, which makes the equilibrium fragile with respect to the addition of dissipation or nonlinearities.  


\subsection{Normal form for the Hamiltonian of the ETG model}

The explicit knowledge of the eigenvalues of the linearized system, makes it possible to cast the Hamiltonian into its normal form, as discussed in Sec.~\ref{ssec:nem}. First of all we point out that the transformation $(\tl_k,\tl_{-k},\tp_k,\tp_{-k})\rightarrow (q_k^1,q_k^2,p_k^1,p_k^2)$, corresponding to
\begin{eqnarray}
q_k^1 &=&\sqrt{\frac{\pi}{k \apc^2}}(\tp_k +\apc \tl_k + \tp_{-k}+\apc \tl_{-k})\,,
\nonumber
\\
p_k^1 &=& -i\sqrt{\frac{\pi}{k \apc^2}}(\tp_k + \apc \tl_k - \tp_{-k}-\apc \tl_{-k})\,,
\nonumber
\\
q_k^2 &=& \sqrt{\frac{\pi}{k }}(\tl_k + \tl_{-k}), \qquad p_k^2 = i\sqrt{\frac{\pi}{k }}(\tl_k - \tl_{-k})\,,
\label{cantr}
\end{eqnarray}
puts the system into canonical Hamiltonian form. Indeed, in terms of the variables $(q_k^1,q_k^2,p_k^1,p_k^2)$, the bracket of (\ref{brk}) takes the canonical form
\beq
\{F,G\}=\sum_{k=1}^{\infty}\frac{\partial F}{\partial q_1^k}\frac{\partial G}{\partial p_1^k}- \frac{\partial F}{\partial p_1^k}\frac{\partial G}{\partial q_1^k}+\frac{\partial F}{\partial q_2^k}\frac{\partial G}{\partial p_2^k}- \frac{\partial F}{\partial p_2^k}\frac{\partial G}{\partial q_2^k}.
\eeq
The Hamiltonian (\ref{hamk}) in terms of the new variables, on the other hand, reads
\beq  \label{hcan}
H_L=\frac{1}{2}\sum_{k=1}^{\infty} \sum_{i,j=1}^4  A_{ij}^k {z_i^k} {z_j^k},
\eeq
where
\beq
A^k= \left( {\begin{array}{cccc}
 a & c & 0 & 0 \\
 c & b & 0 & 0 \\
 0 & 0 & a & -c \\
 0 & 0 & -c & b\\
 \end{array} } \right), \qquad z^k=(q_1^k,q_2^k,p_1^k,p_2^k)
\eeq
with $a=-\rr\apc k$, $b=k(1/(1+k_{\perp}^2)-\rr \apc)$, and $c=\rr|\apc|k$. We emphasize that the transformation (\ref{cantr}) is constructed in such a way that the canonical variables $q_i^k$ and $p_i^k$ are {\it real}. We are thus in the framework depicted in Sec.~\ref{ssec:nem}. For each $k$, the equations of motion are given by
\begin{equation}  
\dot{z^k}=J_c A^k z^k.
\end{equation}
Upon writing the variables as
\beq
q_{1,2}^k=\tilde{z}_{1,2}^k\mathrm{e}^{i\omega^k t}+{{}\tilde{z}_{1,2}^k}^*\mathrm{e}^{-i\omega^k t},\qquad p_{1,2}^k=\tilde{z}_{3,4}^k\mathrm{e}^{i\omega^k t}+{{}\tilde{z}_{3,4}^k}^*\mathrm{e}^{-i\omega^k t},
\eeq
and dropping the tilde in the following, we obtain that the eigenvectors of the linearized system are 
\ben  \label{eigcan}
z_{s}^k=\qs \left( {\begin{array}{c}
 1 \\
 -B_-\\
 -i\\
 -i B_-\\
 \end{array} } \right), \qquad z_{-s}^k={\qs}^* \left( {\begin{array}{c}
 1 \\
 -B_-\\
 i\\
 i B_-\\
 \end{array} } \right), \qquad z_{f}^k=\qf \left( {\begin{array}{c}
 1 \\
 -B_+\\
 -i\\
 -i B_+\\
 \end{array} } \right), \qquad z_{-f}^k={\qf}^* \left( {\begin{array}{c}
 1 \\
 -B_+\\
 i\\
 i B_+\\
 \end{array} } \right).
\een
These are the eigenvectors corresponding to the eigenvalues $\omega_s^k$, $\omega_{-s}^k$, $\omega_f^k$ and $\omega_{-f}^k$, respectively. In (\ref{eigcan}) we introduced the quantities
\beq
B_{\pm}=\frac{b+a\pm\sqrt{(b+a)^2-4c^2}}{2c},
\eeq
whereas ${q_1}_{s,f}^k$ are complex coefficients. Following Sec.~\ref{ssec:nem}, the Lagrange bracket for the slow mode reads
\beq
{z_{-s}^k}^T\Omega z_s^k=2i(1-B_-^2){\qs}^*\qs.
\eeq
Direct calculation shows that
\beq \label{disbm}
1-B_-^2=-\frac{\sqrt{(b+a)^2-4c^2}(\sqrt{(b+a)^2-4c^2}-(b+a))}{2c^2}>0.
\eeq
To obtain the inequality in (\ref{disbm}), we made use of the fact that, for stable modes, $ \sqrt{(b+a)^2-4c^2}<b+a$.\\
The inequality (\ref{disbm}) tells us that, for slow modes, we must pick up the plus sign in the general expression (\ref{lagr}). Moreover, if we choose the normalization constants so that
\beq
\qs={\qs}^*=\frac{1}{D_-}\equiv\frac{1}{\sqrt{1-B_-^2}},
\eeq 
we obtain that the Lagrange bracket for slow modes becomes
\beq
{z_{-s}^k}^T\Omega z_s^k=2i.
\eeq
Of course,
\beq
{z_{s}^k}^T\Omega z_{-s}^k=-2i,
\eeq
and consequently, according to the definition, given in Sec.~\ref{ssec:nem}, we have a PEM, when $\omega_{-s}^k=-\omega_s^k>0$, and a NEM when $\omega_{-s}^k=-\omega_s^k<0$. This confirms the results we obtained in Sec.~\ref{ssec:appetg} with noncanonical variables.

Following the same procedure for the fast mode, we find
\beq
{z_{-f}^k}^T\Omega z_f^k=2i(1-B_+^2){\qf}^*\qf.
\eeq
Given that $1-B_+^2<0$, the Lagrange bracket for the fast mode becomes
\beq  \label{lagrf}
{z_{-f}^k}^T\Omega z_f^k=-2i,
\eeq
after having chosen the following normalization for the eigenvectors: 
\beq
\qf={\qf}^*=\frac{1}{D_+}\equiv\frac{1}{\sqrt{B_+^2-1}}.
\eeq 
Because $\omega_f^k$ is always positive, according to the definition, (\ref{lagrf}) tells us that the fast mode, as expected, is always a PEM.

The transformation that casts the Hamiltonian (\ref{hcan}) into normal form, following Sec.~\ref{ssec:nem}, will be a real canonical transformation $T^k$ that, for each $k$, maps a new set of coordinates $\bar{z}^k=(Q_1^k,Q_2^k,P_1^k,P_2^k)$, in terms of which the Hamiltonian is diagonal, into $z^k$.

After noticing that
\ben
{{z_{-s}^k}^*}^T \Omega z_{-s}^k=-2i,\\
{{z_{f}^k}^*}^T \Omega z_{f}^k=-2i,
\een
the matrix associated with the application $T^k$ is constructed in the following way:
\beq
T^k= \left( {\begin{array}{cccc}
 \frac{1}{D_-} & \frac{1}{D_+} & 0 & 0 \\
 -\frac{B_-}{D_-} & -\frac{B_+}{D_+} & 0 & 0 \\
 0 & 0 & \frac{1}{D_-} & -\frac{1}{D_+} \\
 0 & 0 & \frac{B_-}{D_-} & -\frac{B_+}{D_+}\\
 \end{array} } \right).
\eeq
Direct calculation shows that
\beq
{T^k}^T A^k T^k=\left( {\begin{array}{cccc}
 -\omega_s^k & 0 & 0 & 0 \\
 0 & \omega_f^k & 0 & 0 \\
 0 & 0 & -\omega_s^k & 0 \\
 0 & 0 & 0 & \omega_f^k\\
 \end{array} } \right).
\eeq
Consequently, the Hamiltonian (\ref{hcan}) can be finally written as
\begin{eqnarray} \label{nform}
H_L&=&\frac{1}{2}\sum_{k=1}^{+\infty}\sum_{i,j=1}^4{{(T^k\bar{z}^k)}_i}^T A_{ij}^k{(T^k\bar{z}^k)}_j
\nonumber\\
&=&
\frac{1}{2} \sum_{k=1}^{+\infty}\omega_f^k\left({{Q_2^k}^2+{P_2^k}^2}\right)-{\omega_s^k}\left({{Q_1^k}^2 +{P_1^k}^2}\right).
\end{eqnarray}
The expression (\ref{nform}) corresponds to the normal form for the Hamiltonian of the linearized ETG model for stable modes. It clearly shows how the corresponding energy can be decomposed into the sum of energies of harmonic oscillators which possess, as characteristic frequencies, those of the fast and slow modes. The harmonic oscillators associated with the fast modes always provide a positive contribution to the total energy. Those associated to the slow modes, on the other hand, give a negative contribution if $\omega_s^k>0$, which translates into the condition on the equilibrium pressure gradient discussed in Sec.~\ref{ssec:appetg}.


\section{Conclusions} \label{sec:concl}

By making use of the Hamiltonian formalism, we have analyzed the mode signature and the stability properties of an ETG fluid model. The families of Casimir invariants of the model were obtained, thereby showing that the dynamics of the model is subject to an infinite number of constraints. A  stability condition has been derived, according to which, the absence of equilibrium flow and a restriction on the pressure equilibrium profile imply stability. Subsequently, after reviewing the concept of mode signature in the Hamiltonian framework, we have explicitly determined the energies of stable modes. From the stability viewpoint, the dispersion relation gives us a spectral stability condition which, however, does not give us information about the stronger condition of  energy stability. Indeed, our analysis shows that spectrally stable homogeneous equilibria can be of two types, depending on the value of the parameters. If $\apc <0$, equilibria are spectrally and energy stable (i.e. with no NEMs). If $0<\apc<1/(4(1+k_{\perp}^2)\rr)$, on the other hand, equilibria are still spectrally  stable, but they are not energy stable. Indeed, the sign of the second variation of the free energy functional, in this case, is indefinite because of the presence of NEMs. Equilibria of the second type might then be prone to dissipation-induced or nonlinearity-induced instabilities.

As anticipated in Sec.~\ref{sec:intro}, one of the advantages of the Hamiltonian formalism for investigating stability and mode signature, is that it is very general and can be applied to any plasma model with a Hamiltonian structure. The method described and applied in the present paper, nevertheless, refers only to systems with discrete spectrum. A natural and promising future project, would be to analyze mode signature and perform the normal form analysis for more complex equilibria that support  continuous spectrum using the techniques developed in \cite{Mor92,Mor00,HagMor10,Hir10}.  Such equilibria  have more spatial dependence, for example.

\acknowledgments
E. Tassi acknowledges fruitful discussions with \"{O}. D. G\"{u}rcan during the Festival de Th\'eorie 2009. It is also a pleasure to acknowledge useful discussions with  the Nonlinear Dynamics group at the Centre de Physique Th\'eorique, Luminy. This work was supported by the European Community under the contracts of Association between EURATOM, CEA, and the French Research Federation for fusion studies. The views and opinions expressed herein do not necessarily reflect those of the European Commission. Financial support was also received from the Agence Nationale de la Recherche (ANR EGYPT). In addition, PJM  was supported  by the US Department of Energy Contract No.~DE-FG03-96ER-54346.

\end{document}